\documentclass[arxiv,usenatbib]{iupartex}

\usepackage{newtxtext,newtxmath}
\usepackage[T1]{fontenc}
\usepackage{natbib}

\usepackage{adjustbox} 
\usepackage[capitalise]{cleveref}

\title{A Systematic Replicability and Comparative Study of BSARec and SASRec for Sequential Recommendation}

\author[D'Ercoli et. al]{%
Chiara D'Ercoli $^{1}$,\orcid{0009-0005-5334-8400}
Giulia Di Teodoro $^{2}$,\orcid{0000-0003-1339-6983}
Federico Siciliano$^{1}$\orcid{0000-0002-0418-0067}
 \\
 \\
$^1$ La Sapienza University, Department of Computer, Automatic and Management Engineering Antonio-Ruberti (DIAG)\\
$^2$ University of Pisa, Information Engineering Department\\
} 

\pubyear{2025}

\begin{document}
\label{firstpage}
\pagerange{\pageref*{firstpage}--\pageref*{lastpage}}
\maketitle

\begin{abstract}
This study aims at comparing two sequential recommender systems: Self-Attention based Sequential Recommendation (SASRec), and Beyond Self-Attention based Sequential Recommendation (BSARec) in order to check the improvement frequency enhancement - the added element in BSARec - has on recommendations. The models in the study, have been re-implemented with a common base-structure from EasyRec, with the aim of obtaining a fair and reproducible comparison. The results obtained displayed how BSARec, by including bias terms for frequency enhancement, does indeed outperform SASRec, although the increases in performance obtained, are not as high as those presented by the authors. This work aims at offering an overview on existing methods, and most importantly at underlying the importance of implementation details for performance comparison. 

\end{abstract}




\section{Introduction}
In the field of Recommender Systems (RecSys), topics such as sequential recommendation and next-item prediction have become increasingly popular. Sequential recommendation refers to systems that aim at modeling users’ evolving preferences over time, namely by capturing their preferences from historical interactions and using them to predict the next item they will interact with \cite{seqRecSys}. Sequential RecSys represent an improvement from normal RecSys, as they have the added ability to capture the temporal order of interactions of a user, therefore enabling the collection of their dynamic preferences in a more accurate manner. 
Given their potential, recent work has focused on the development of sequential RecSys, obtaining outstanding results and capturing several interesting characteristics. Most of these models are transformer-based, hence they utilize architectures that were originally built for natural language processing (NLP) tasks, and examples are Bidirectional Encoder Representations from Transformers for sequential Recommendation (BERT4Rec) \cite{bert4rec}, Self-Attention based Sequential Recommendation (SASRec) \cite{sasrec}, and Beyond Self-Attention based Sequential Recommendation (BSARec) \cite{bsarec}. 

Recent work has raised concerns about the reliability of evaluation protocols in Sequential RecSys. In particular, \citeauthor{goldLoss}\cite{goldLoss} highlight that performance comparisons across models can be misleading when based on inconsistent experimental conditions—such as differing loss functions, mismatched hyperparameters, or implementation-specific choices. Such inconsistencies are not uncommon in the literature, where some studies report superior performance for specific models without a fully aligned evaluation setup \cite{gsasrec}.
Such claims stem from some works reimplementing existing architectures "from scratch", rather than building on reilable packages and libraries (i.e., PyTorch - e.g., \texttt{torch.nn}; see \cite{pytorch}).  
To ensure generalization and reproducibility, implementations of existing architectures and frameworks should, where possible, rely on native PyTorch components. This facilitates transparent reporting of experimental settings, allowing the research community not only to fully understand and adapt the proposed analyses, but also to accurately assess model performance and make informed comparisons when selecting models for specific tasks.

Our aim is to shed light on model replicability and evaluation in the domain of sequential RecSys. In particular, we focus on two influential models: Self-Attention based Sequential Recommendation (SASRec) \cite{sasrec}, a model that leverages the Transformer’s self-attention mechanism to capture users’ dynamic preferences over time, and Beyond Self-Attention for Sequential Recommendation (BSARec) \cite{bsarec}, which addresses a key limitation of self-attention known as the oversmoothing problem, where hidden representations become overly similar, hindering the model’s ability to distinguish fine-grained sequential patterns. 
By providing a rigorous and fair comparison based on the exact same base implementation, we seek to verify whether BSARec truly outperforms SASRec as claimed.

\section{Related Work}

Sequential RecSys aim at recommending the next item the user might interact with based on their own historical interactions. 

Several notable works have been conducted in this area,
each presenting distinct approaches. Although many techniques have tried employing Markov Chains \cite{he2016markov} or Convolutional Neural Networks (CNN) \cite{dl}
for sequence modeling, recently, the advancements in deep neural network-based recommendation algorithms has influenced research towards: recursion - as done in GRU4Rec \cite{gru4rec}, where recurrent neural networks are employed to provide session-based recommendations, self-attention mechanisms - as in SASRec \cite{sasrec}, and finally, towards transformers, as done in BERT4Rec \cite{bert4rec}, where the bidirectionality is exploited for more reliable historical representations. 

These models and techniques all aim at reliably capturing preferences the user has expressed in the past. An issue more recent systems have tried to solve regards the characteristic of transformer-based RecSys to focus on lower frequencies when recommending items to the user, failing to consider also higher frequencies - all due to the low-pass filtering nature of self-attention. DuoRec's \cite{duorec} goal was to improve this aspect by employing contrastive learning, but an even more interesting approach was presented in FEARec \cite{fearec}, where the model uses time domain attention and autocorrelation, with frequency enhancement. 
All these models aimed at solving the low-pass filter nature of RecSys and their inability to capture high-frequency information and to distinguish the inherent periodicity obscured in the time domain.
The models compared in this study, had the same goal: SASRec \cite{sasrec} - employing self-attention, and its extention BSARec \cite{bsarec} - including an inductive bias. 

\section{Methodology }
In this section, the chosen models to compare will be presented in more detail. 

\textbf{SASRec} \cite{sasrec} is a sequential recommendation model that introduces the self-attention mechanisms from the Transformer's architecture \cite{bert} into the recommendation domain. 

The architecture consists of an input embedding layer, where each item in the sequence is mapped to a dense vector representation. To preserve the order of interactions, a positional embedding is added to each item embedding. The resulting sequence of embeddings is then passed to one or more Transformer blocks, each composed of  a multi-head self-attention mechanism followed by a pointwise feed-forward network (FFN). After both the attention and feed-forward sublayers, layer normalization and residual connections are applied to facilitate training stability and improve convergence. The self-attention mechanism allows the model to learn which past items in the sequence are more relevant for predicting future interactions, while causal masking ensures that the model only attends to previous positions in the sequence, preventing information leakage from future items. During training, the model predicts the next item at each position in the sequence by computing a relevance score through the dot product between the hidden state at that position and the embeddings of candidate items. At inference time, only the final hidden state (corresponding to the last observed interaction) is used to predict the next likely item.

\textbf{BSARec} \cite{bsarec} is introduced to address the lack of inductive bias in transformer-based RecSys models, which often focus on long-range dependencies while neglecting fine-grained sequential patterns.This limitation is attributed to the low-pass filtering nature of self-attention mechanism,  which tends to oversmooth item embeddings by emphasizing global signals and suppressing high-frequency signals that reflect short-term behaviors.
To overcome this limitation, BSARec extends the work in \cite{sasrec} by introducing the BSALayer that captures both low frequency (long-term) and high frequency (short-term) dependencies in user behavior sequences. 
The core innovation lies in modeling the attention mechanism in the frequency domain: the authors apply the Discrete Fourier Transform (DFT) to the self-attention matrix and design an inductive bias component that acts as a high-pass filter. This component amplifies high-frequency signals, allowing the model to retain important short-term patterns that would otherwise be smoothed out.

The difference between SASRec and BSARec lies in the block between the embedding layer and the FFN. While SASRec applies a self-attention layer followed by normalization and the point-wise FFN, BSARec introduces a parallel structure: it applies a BSALayer, which consists of the standard self-attention mechanism, alongside an attentive inductive bias module that incorporates frequency re-scaling. Both components are independently normalized before being combined.
The Fast Fourier Transform (FFT) is applied in the attentive inductive bias module, where from the input signal, low and high filters are extracted according to the hyperparameter $c$, representing the cutoff frequency that separates low-frequency components from high-frequency ones. These are then recombined through an inverse FFT, weighted by a learnable parameter $\beta$ that controls the contribution of each frequency component— hence the term frequency re-scaling. These two - the self-attention and the attentive and inductive bias blocks - are added considering a parameter $\alpha$, representing the weight to use for the consideration of the inductive bias.

\begin{figure}
    \centering
    \includegraphics[width=0.5\linewidth]{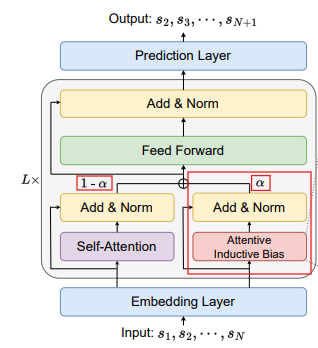}
    \caption{BSARec's architecture. Adapted from \protect\cite{bsarec}. The blocks highlighted in red represent the added part to SASRec's architecture.}
    \label{fig:bsarec_sasrec}
\end{figure}

In order to foster reproducibility, clarity and efficiency, the EasyRec \cite{fedesrepo} library has been utilized for the comparative study developments.  
EasyRec is a versatile Python library designed to streamline the process of configuring and building Sequential RecSys models while exploiting PyTorch Lightning and PyTorch models.The goal of this library is to standardize data preprocessing and model implementations and establishing a
foundation for consistent and reproducible experimentation. Here, YAML files are used to include experiment configurations, model hyperparameters, training and data elements. A main Python script reads such configurations and streamlines model training.  
Both SASRec and BSARec have been re-implemented through this library, so as to ensure maximum comparability. 

SASRec's implementation inherits from \texttt{torch.nn.Module}, and its structure adheres to the original implementation in \cite{sasrec} but uses all PyTorch-native components. As a matter of fact, the difference between EasyRec's \cite{fedesrepo} and the original paper's\cite{sasrec} implementation, only lies in the latter's manual reimplementation of all building blocks of the model, including the multihead attention mechanism and the embedding layer. In contrast, SASRec implementation in EasyRecleverages built-in PyTorch modules such as \texttt{Embedding}, \texttt{Dropout}, \texttt{TransformerEncoderLayer}, \texttt{TransformerEncoder}, and \texttt{LayerNorm} to build the model architecture more modularly and efficiently. 

BSARec's implementation in EasyRec was similarly developed, although it required less generalizable steps, as - given the addition of some specific parameters - a more tailored implementation was necessary. 
Built on top of EasyRec, BSARec inherits from SASRec and uses PyTorch's \texttt{TransformerEncoder} module along with a custom-defined encoder layer called \texttt{BSARecEncoderLayer}. This encoder layer includes the parameter $\alpha$, and the custom "filter" layer (called \texttt{BSARecLayer} - different from the BSALayer, which is instead, the conceptual union of the filter block and self-attention attention block). In this model, the \texttt{BSARecEncoderLayer} inherits from the \texttt{torch.nn.TransformerEncoderLayer} module and extends it with the inclusion of the filter layer and its actual application - all according to the parameter $\alpha$. The encoder layer in fact, after calling the filter layer - where dropout and layer normalization (both torch modules) are applied, and where the Fourier Transform is calculated according to different parameters ($\beta$ and c) - returns the sequence of embeddings. More specifically, both the Fourier Transform and the dropout layer are applied, and to this output, by adding the input to a self-attention block, the final result is calculated. This output is the result of blending the initial output and the attention output according to $\alpha$, passing it through a feedforward block and through a final residual connection (normalization in this case is applied either before or after).

\subsection{Experimental setup}
\subsubsection{Datasets \& Metrics}
In this study, two data sets are used: Movielens 1m (ml-1m) \cite{movielens} and Foursquare-nyc (fs-nyc) \cite{foursquare}. Ml-1m, consists of 3,883 nodes representing movies and 6040 nodes representing users, with approximately 1 million ratings between the two, with ratings represented as ground truth labels. Fs-nyc, instead, includes check-in data from New York City (NYC) and Tokyo, collected from April 12, 2012 to February 16, 2013. It contains 227,428 NYC check-ins, each associated with its time stamp, GPS coordinates and semantic information such as venue-categories. 

The top-k ranking metrics considered for the evaluation are widely used in the field of RecSys, and they are: Normalized Discounted Cumulative Gain@k (NDCG@k) with $k\in\{5,10,20\}$ , Precision@10, and Recall@10. 

\subsubsection{Hyperparameters tuning}
The analysis considered the same parameters presented in the respective models' papers, whereby tuning was performed only on $\alpha$ and c - for BSARec, and the dropout rate for SASRec. In the first case, the ranges considered were: 
\begin{itemize}
    \item $\alpha$ = [0.1, 0.5, 0.7, 0.9]
    \item c = [1, 3, 5, 7, 9]
    \item dropout rate = [0.0005, 0.2]
\end{itemize}
The best results for BSARec are obtained with $\alpha$ = 0.7 and c = 1 (for ml-1m) - which respects what found by \citeauthor{bsarec}, and $\alpha$ = 0.3 and c = 1 (for fs-nyc). For SASRec instead, the best performing model is obtained when the dropout rate is 0.0005 for both datasets.

\section{Results} 
The results obtained are presented in Table \ref{metrics_table}.

From the metrics, it emerges that BSARec's performance is higher than SASRec's. On  ml-1m, BSARec obtains both higher precision@10 and recall@10. The same is visible for NDCG@k (k = 5,10,20) values, where performance on BSARec increases incrementally. This dataset was also explored in the original work \cite{bsarec}, but the results obtained in this comparative study are slightly lower than the original ones. This might suggest that implementation details might have impacted \citeauthor{bsarec}'s scores. 

A wider difference in performance is visible with the fs-nyc dataset, where, aside from NDCG@5 scores - BSARec performs evidently higher than SASRec. This might be due to the geographic nature of fs-nyc nature of data, as the temporal aspect is more impactful - hence supporting the frequency identification and filtering performed by BSARec. Gains are larger on the foursquare-nyc dataset - ranging from 10.3\% to 14.3\% - while on ml-1m they are more modest, between 3.7\% and 5.7\%. 

BSARec indeed succeeds in filling the gap left by the inherent inductive bias and low-pass nature stemming from transformers' self-attention mechanism. BSARec's vanilla self-attention and frequency-oriented mechanism, enables the model to capture a combination of inductive bias and does indeed mitigate oversmoothing.

\begin{table}[h]
\centering
\begin{adjustbox}{width=\textwidth}
\begin{tabular}{llllllllllll}
\toprule
Model & Dataset &  NDCG@5 & NDCG@10 &  NDCG@20 & Precision@10 & Recall@10 \\
\midrule
BSARec & ml-1m & \textbf{0.06172} & \textbf{0.07704} & \textbf{0.0919}  & \textbf{0.0134} & \textbf{0.1341} \\
SASRec & ml-1m  & 0.05854 &	0.0729	& 0.0886  & 0.0127 & 0.1275 \\
BSARec & foursquare-nyc  & \textbf{0.2668}	& \textbf{0.2746}	& \textbf{0.2819} & \textbf{0.0311} & \textbf{0.3112} \\
SASRec & foursquare-nyc  & 0.2334	& 0.2421	& 0.2475 & 0.0282 & 0.2816 \\
\bottomrule
\end{tabular}
\end{adjustbox}
\caption{Metrics obtained for ml1m and fs-nyc for both models BSARec and SASRec. The best metric for each dataset is highilighted in bold. }
\label{metrics_table}
\end{table}

\begin{figure}[!h]
    \centering
    \begin{minipage}{0.43\linewidth}
        \centering
        \includegraphics[width=\linewidth]{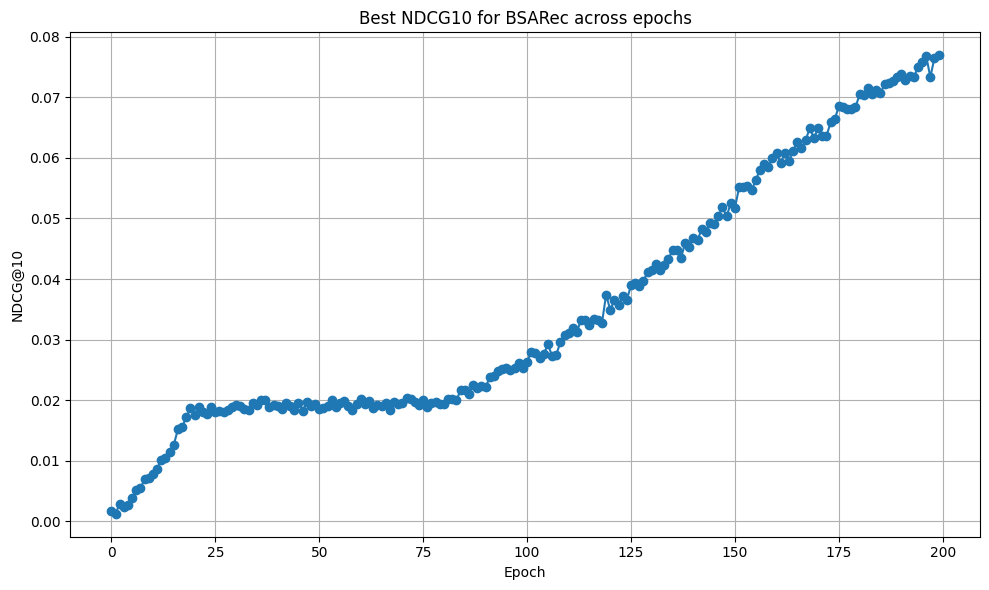}
        \caption{BSARec}
        \label{fig:best_bsarec}
    \end{minipage}%
    \hfill
    \begin{minipage}{0.43\linewidth}
        \centering
        \includegraphics[width=\linewidth]{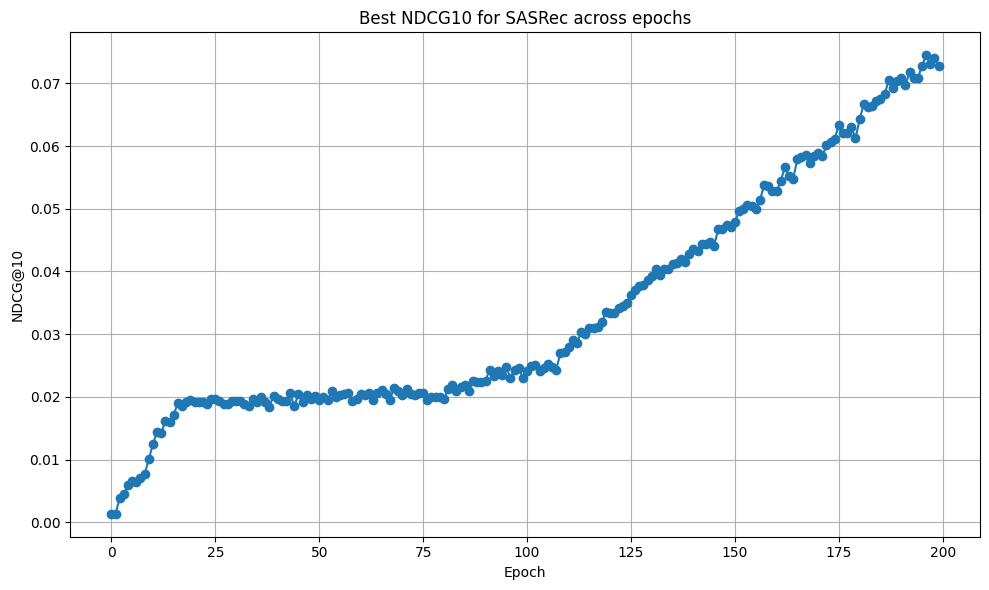}
        \caption{SASRec}
        \label{fig:best_sasrec}
    \end{minipage}%
    \hfill
    \caption{BSARec (\cref{fig:best_bsarec}) and SASRec's (\cref{fig:best_sasrec}) NDCG@10 reported across training epochs.}
    \label{fig:best_models}
\end{figure}

In Figure \ref{fig:best_models} the two best performing models are reported, namely, their NDCG@10 scores across epochs are. It is evident how the model immediately improves its performance after 100 epochs, and how BSARec's scores tend to be slightly higher than SASRec's. No extreme oscillations are visible.

\section{Conclusion}
This comparative study aimed at presenting the added value brought by the authors in \citeauthor{bsarec} to sequential recommender systems - namely SASRec \cite{sasrec} - in terms of frequency enhancement. The goal of the study was to present a more trustworthy and unbiased comparison between the two models, re-implementing them using the same base architecture and only adding the additional parameters and layers for the inclusion of higher frequency terms - and the consequent adjustment of transformers' inherent low-pass filtering nature. 

The results recorded in this comparative study do confirm original BSARec's scores, although revealing a slight decrease in performance, highlighting the importance of model implementation alignment when presenting methodologies that extend an existing model. 
This study paves the way for future research on the topic of enhanced attention modules for sequential recommendation and inductive bias augmentation.

\bsp	

\label{lastpage}

\begin{thebibliography}{16}
\providecommand{\natexlab}[1]{#1}
\providecommand{\url}[1]{\texttt{#1}}
\expandafter\ifx\csname urlstyle\endcsname\relax
  \providecommand{\doi}[1]{doi: #1}\else
  \providecommand{\doi}{doi: \begingroup \urlstyle{rm}\Url}\fi

\bibitem[Betello et~al.(2024)Betello, Purificato, Siciliano, Trappolini, Bacciu, Tonellotto, and Silvestri]{fedesrepo}
Filippo Betello, Antonio Purificato, Federico Siciliano, Giovanni Trappolini, Andrea Bacciu, Nicola Tonellotto, and Fabrizio Silvestri.
\newblock A reproducible analysis of sequential recommender systems.
\newblock \emph{IEEE Access}, 2024.

\bibitem[Boka et~al.(2024)Boka, Niu, and Neupane]{seqRecSys}
Tesfaye~Fenta Boka, Zhendong Niu, and Rama~Bastola Neupane.
\newblock A survey of sequential recommendation systems: Techniques, evaluation, and future directions.
\newblock \emph{Information Systems}, 125:\penalty0 102427, 2024.
\newblock ISSN 0306-4379.
\newblock \doi{https://doi.org/10.1016/j.is.2024.102427}.
\newblock URL \url{https://www.sciencedirect.com/science/article/pii/S0306437924000851}.

\bibitem[Du et~al.(2023)Du, Yuan, Zhao, Qu, Zhuang, Liu, and Sheng]{fearec}
Xinyu Du, Huanhuan Yuan, Pengpeng Zhao, Jianfeng Qu, Fuzhen Zhuang, Guanfeng Liu, and Victor~S. Sheng.
\newblock Frequency enhanced hybrid attention network for sequential recommendation, 2023.
\newblock URL \url{https://arxiv.org/abs/2304.09184}.

\bibitem[Harper and Konstan(2015)]{movielens}
F~Maxwell Harper and Joseph~A Konstan.
\newblock The movielens datasets: History and context.
\newblock \emph{Acm transactions on interactive intelligent systems (tiis)}, 5\penalty0 (4):\penalty0 1--19, 2015.

\bibitem[He and McAuley(2016)]{he2016markov}
Ruining He and Julian McAuley.
\newblock Fusing similarity models with markov chains for sparse sequential recommendation.
\newblock In \emph{2016 IEEE 16th international conference on data mining (ICDM)}, pages 191--200. IEEE, 2016.

\bibitem[Hidasi et~al.(2016)Hidasi, Karatzoglou, Baltrunas, and Tikk]{gru4rec}
Balázs Hidasi, Alexandros Karatzoglou, Linas Baltrunas, and Domonkos Tikk.
\newblock Session-based recommendations with recurrent neural networks, 2016.
\newblock URL \url{https://arxiv.org/abs/1511.06939}.

\bibitem[Kang and McAuley(2018)]{sasrec}
Wang-Cheng Kang and Julian McAuley.
\newblock Self-attentive sequential recommendation, 2018.
\newblock URL \url{https://arxiv.org/abs/1808.09781}.

\bibitem[Klenitskiy and Vasilev(2023)]{goldLoss}
Anton Klenitskiy and Alexey Vasilev.
\newblock Turning dross into gold loss: is bert4rec really better than sasrec?
\newblock In \emph{Proceedings of the 17th ACM Conference on Recommender Systems}, RecSys '23, page 1120–1125, New York, NY, USA, 2023. Association for Computing Machinery.
\newblock ISBN 9798400702419.
\newblock \doi{10.1145/3604915.3610644}.
\newblock URL \url{https://doi.org/10.1145/3604915.3610644}.

\bibitem[LeCun et~al.(2015)LeCun, Bengio, and Hinton]{dl}
Yann LeCun, Y.~Bengio, and Geoffrey Hinton.
\newblock Deep learning.
\newblock \emph{Nature}, 521:\penalty0 436--44, 05 2015.
\newblock \doi{10.1038/nature14539}.

\bibitem[Paszke et~al.(2019)Paszke, Gross, Massa, Lerer, Bradbury, Chanan, Killeen, Lin, Gimelshein, Antiga, Desmaison, Köpf, Yang, DeVito, Raison, Tejani, Chilamkurthy, Steiner, Fang, Bai, and Chintala]{pytorch}
Adam Paszke, Sam Gross, Francisco Massa, Adam Lerer, James Bradbury, Gregory Chanan, Trevor Killeen, Zeming Lin, Natalia Gimelshein, Luca Antiga, Alban Desmaison, Andreas Köpf, Edward Yang, Zach DeVito, Martin Raison, Alykhan Tejani, Sasank Chilamkurthy, Benoit Steiner, Lu~Fang, Junjie Bai, and Soumith Chintala.
\newblock Pytorch: An imperative style, high-performance deep learning library, 2019.
\newblock URL \url{https://arxiv.org/abs/1912.01703}.

\bibitem[Petrov and Macdonald(2023)]{gsasrec}
Aleksandr~Vladimirovich Petrov and Craig Macdonald.
\newblock gsasrec: Reducing overconfidence in sequential recommendation trained with negative sampling.
\newblock In \emph{Proceedings of the 17th ACM Conference on Recommender Systems}, RecSys ’23, page 116–128. ACM, September 2023.
\newblock \doi{10.1145/3604915.3608783}.
\newblock URL \url{http://dx.doi.org/10.1145/3604915.3608783}.

\bibitem[Qiu et~al.(2022)Qiu, Huang, Yin, and Wang]{duorec}
Ruihong Qiu, Zi~Huang, Hongzhi Yin, and Zijian Wang.
\newblock Contrastive learning for representation degeneration problem in sequential recommendation.
\newblock In \emph{Proceedings of the Fifteenth ACM International Conference on Web Search and Data Mining}, WSDM ’22, page 813–823. ACM, February 2022.
\newblock \doi{10.1145/3488560.3498433}.
\newblock URL \url{http://dx.doi.org/10.1145/3488560.3498433}.

\bibitem[Shin et~al.(2024)Shin, Choi, Wi, and Park]{bsarec}
Yehjin Shin, Jeongwhan Choi, Hyowon Wi, and Noseong Park.
\newblock An attentive inductive bias for sequential recommendation beyond the self-attention, 2024.
\newblock URL \url{https://arxiv.org/abs/2312.10325}.

\bibitem[Sun et~al.(2019)Sun, Liu, Wu, Pei, Lin, Ou, and Jiang]{bert4rec}
Fei Sun, Jun Liu, Jian Wu, Changhua Pei, Xiao Lin, Wenwu Ou, and Peng Jiang.
\newblock Bert4rec: Sequential recommendation with bidirectional encoder representations from transformer, 2019.
\newblock URL \url{https://arxiv.org/abs/1904.06690}.

\bibitem[Vaswani et~al.(2017)Vaswani, Shazeer, Parmar, Uszkoreit, Jones, Gomez, Kaiser, and Polosukhin]{bert}
Ashish Vaswani, Noam Shazeer, Niki Parmar, Jakob Uszkoreit, Llion Jones, Aidan~N. Gomez, \L{}ukasz Kaiser, and Illia Polosukhin.
\newblock Attention is all you need.
\newblock In \emph{Proceedings of the 31st International Conference on Neural Information Processing Systems}, NIPS'17, page 6000–6010, Red Hook, NY, USA, 2017. Curran Associates Inc.
\newblock ISBN 9781510860964.

\bibitem[Yang et~al.(2015)Yang, Zhang, Zheng, and Yu]{foursquare}
Dingqi Yang, Daqing Zhang, Vincent Zheng, and Zhiyong Yu.
\newblock Modeling user activity preference by leveraging user spatial temporal characteristics in lbsns.
\newblock \emph{Systems, Man, and Cybernetics: Systems, IEEE Transactions on}, 45:\penalty0 129--142, 01 2015.
\newblock \doi{10.1109/TSMC.2014.2327053}.

\end{thebibliography}
\end{document}